\documentclass[12pt]{article}
\usepackage{float}
\usepackage{bm}
\usepackage{dsfont}
\usepackage{fancybox}
\usepackage [pdftex,linkcolor=blue]{hyperref}
\usepackage{varioref}
\usepackage {multicol}
\usepackage{color}
\usepackage{graphicx} 
\usepackage{amssymb}
\usepackage{amsmath}
\usepackage{mathtools}
\usepackage{enumerate}
\usepackage{cleveref}
\usepackage{setspace}
\usepackage{url}
\let\oldurl\url
\usepackage{hyperref}

\let\url\oldurl
\usepackage{amsfonts}
\usepackage{dsfont}
\usepackage{upgreek}
\usepackage{titlesec}
\usepackage[sort&compress]{natbib}

\title{PLS Generalized Linear Regression and Kernel Multilogit Algorithm (KMA) for Microarray Data Classification}
\author{Adolphus Wagala,
Graciela Gonz\'alez-Far\'ias,
 Rogelio Ramos,\\ and Oscar Dalmau \\
\vspace{6pt} Centro de Investigaci\'on en Matem\'aticas A.C.,\\Jalisco S/N, Col. Valenciana,\\CP: 36240,Guanajuato, Gto, M\'exico.\\}

\date{}

\begin{document}
\maketitle

\begin{abstract}
 We implement extensions of the partial least squares generalized linear regression (PLSGLR)due to \cite{Bastien2005} through its combination with logistic regression and linear discriminant analysis, to get a partial least squares generalized linear regression-logistic regression model (PLSGLR-log), and a partial least squares generalized linear regression-linear discriminant analysis model (PLSGLRDA). These two classification methods are then compared with classical methodologies like the k-nearest neighbours  (KNN), linear discriminant analysis (LDA), partial least squares discriminant analysis (PLSDA), ridge partial least squares (RPLS), and support vector machines(SVM). Furthermore, we implement the kernel multilogit algorithm (KMA) by \cite{Dalmau2015} and compare its performance with that of the other classifiers. The results indicate that for both un-preprocessed and preprocessed data, the KMA has the lowest classification error rates. 

\end{abstract}

\section{Introduction}
 	The field of genomics has witnessed a tremendous increase in the amount of data generation due to biotechnological advances like microarrays and next-generation sequencing platforms. These biotechnological advances have made it possible to simultaneously monitor expression levels for thousands of genes, and thus help in solving particular problems related to the identification of molecular variants in them, and their relation to the classification, diagnosis, prognosis and treatment of different conditions. The high dimensional data generated from microarray technology involve many thousands of genes measured simultaneously, a different microarray for each individual. This definitely introduces some noise and unwanted variations that might stem from technical or unknown sources.
 
 In a microarray experiment let  $n$ and  $p$ be the numbers of the samples and genes respectively, so that the generated data is a $n \times p$ matrix. The main challenge with these technologies is that the resultant data are noisy due to biological and technological variations, and at the same time they usually are high dimensional, i.e., they have more variables than cases due to a low sample size, so  $n<<p$. This condition makes the direct application of most classical statistical methodology implausible, leading researchers to propose new solutions for this type of problem. 
 
 Normally before the down stream analysis of the data generated from DNA microarrays, a preprocessing and normalization stage is performed to remove the noise, filtering out the genes with low expression values, addressing missing values, and standardizing the data via a log-transformation. One of the most used preprocessing procedures for microarray data was proposed by \cite{Dudoitetal2002}, which entails three basic steps, namely: thresholding, filtering out of genes outside of a range of minimum/maximum intensities, and finally, standardization of the expression values by a log transformation (see(\cite{Alshamlan2013, Dudoitetal2002})). 
 
 This work considers classification problems for microarray data sets under two conditions: un-preprocessed and preprocessed. In the un-preprocessed data all genes available in the study are included, while in the preprocessed only the subset of genes believed to play important roles in the biological problem of interest are used. We extend the Partial Least Squares Generalized Linear Regression (PLSGLR) algorithm of \cite{Bastien2005} by combining it with Logistic Regression, to give PLSGLR-log, and with Linear Discriminant Analysis to come up with PLSGLRDA. Furthermore, we compare their performance with that of the kernel multilogit algorithm (KMA) proposed by \citep{Dalmau2015}, and of the classical methods: the k-Nearest Neighbour (KNN), Ridge Partial Least Squares (RPLS), Partial Least Squares-Linear Discriminant Analysis (PLSDA), the usual Linear Discriminant Analysis (LDA) and the Support Vector Machines (SVM), when applied to a set of microarray data, referred to in this work as the Colon data set (by \citep{Alon1999}). We evaluate the classifiers with regard to their classification error rates in this data set and compare them.
 
 Our work addresses problems similar to many studies involving classification in microarrays, with typically high dimensional data and low numbers of samples (or subjects). Following a two stage strategy, many involve the use of the original PLS to build the components, even though the response variables are discrete, for example the analysis of \citep{Nguyen2002b,Nguyen2002a}; this is intuitively not correct since the original PLS is an algorithm best suited for continuous response variables.  And in almost all of the procedures a variable (gene) selection step is implemented, with an accompanying computing cost.  This paper describes a procedure suitable for categorical data, and its performance is studied with and without the gene selection step, and compared to that of each of the other classifiers used. An additional advantage of our approach is that the PLSGLR can deal with missing values, unlike the original PLS, commonly used in the literature.
 
 The proposed two stage strategy for the classification problem is described as follows.
 \begin{table}[H]
 	\caption{\label{Strategy1} Proposed strategy}
 	\vspace{.5cm}
 	\begin{center}
 		\small
 		\begin{tabular}{l}
 			\hline
 			\textbf{Steps}\\
 			\hline \hline \\
 			\begin{minipage}{5in}
 				
 				\textbf{Step 1: Dimension reduction} \\
 				In this stage, we propose to use PLSGLR to project the high dimensional data to a low dimension space thus resulting in new components (latent variables), which preserve the information in the intrinsic structure of the data. \\
 				
 				\textbf{Step 2: Use of latent variables for classification} \\
 				Analyze the obtained latent variables with the classical statistical classifiers:
 				\begin{enumerate}[i]
 					\item  PLSGLR components with logistics regression to get the PLSGLR-logistic model denoted as (PLSGLR-log)
 					\item PLSGLR components with linear discriminant analysis model to get PLSGLR-Linear Discriminant Analysis model denoted as (PLSGLRDA)
 				\end{enumerate}
 				
 			\end{minipage}
 			\\
 			\\
 			\hline
 		\end{tabular}
 	\end{center}
 \end{table}
 To the best of our knowledge, the proposed combination of PLS generalized linear regression algorithm with logistic and discriminant analysis has not been used before in cases where $n<<p$. The PLS generalized linear regression algorithm is simple, and a good performance when compred to the classical methods would make it an attractive alternative.
 
\section{Kernel multilogit algorithm (KMA)}
The KMA was recently proposed by \cite{Dalmau2015}. This algorithm works by first transforming a categorical response variable to a continuous one via a multilogit transformation. The transformed variable is then used with the explanatory variables in a regression model for classification and prediction. Finally, the new predicted variables are transformed back using the inverse multilogit function to the original space to enable classification. 

Let the response variable vector $\bm{y}$  be categorical with class labels $\{1,2,\ldots,C\}$.  To classify a discrete variable from predictor variables $\bm{x}$, the first step is to transform the response variable $\bm{y}$ into a new space using the multilogit function. The multinomial logit model with $C$ as the reference category can be given as 
\begin{align}\label{Multinom}
\begin{split}
\text{Pr}(\bm{y}&=j|\bm{x})=\frac{\text{exp}\{f(\bm{x};\bm{\theta}_j)\}}{1+\sum\limits_{i=1}^{C-1}\text{exp}\{f(\bm{x};\bm{\theta}_i)\}},\quad j=\{1,2,\ldots,C-1\} \\
\text{Pr}(\bm{y}&=C|\bm{x})=\frac{1}{1+\sum\limits_{i=1}^{C-1}\text{exp}\{f(\bm{x};\bm{\theta}_i)\}},
\end{split}
\end{align}
where $f(\bm{x};\bm{\theta}_i)=\bm{x}^T\bm{\theta}_i$.
The expected value of $\bm{y}$ being a multinomial random variable is given by  $E(\bm{y}|\bm{x})=[\text{Pr}(\bm{y}=1|\bm{x}),\text{Pr}(\bm{y}=2|\bm{x}),\ldots,\text{Pr}(\bm{y}=C|\bm{x})]^T$. Now, denoting $\bm{t}=E(\bm{y}|\bm{x})$, the original response variable $\bm{y}$ is not used but instead a transformed version $\bm{\vartheta}=\text{logit}(\bm{t})$ is used. The logit transformation is done with $C$ as the reference category as follows 
\begin{align}\label{Logit}
\vartheta_j=\text{logit}(t_j)=\text{log}\frac{t_j}{t_C},\quad j=\{1,2,\ldots,C-1\} 
\end{align}
where $\vartheta_j\in \bm{\vartheta}, t_j\in \bm{t}$. 

In the second step  a parametric linear model is proposed and its parameter estimates can be obtained via the standard Bayesian formula $\text{Pr}(\bm{\vartheta}|\bm{x})=\text{Pr}(\bm{x}|\bm{\vartheta})\text{Pr}(\bm{\vartheta})/\text{Pr}(\bm{x})$  where $\text{Pr}(\bm{\vartheta}|\bm{x})$ is the posterior probability distribution, $\text{Pr}(\bm{x}|\bm{\vartheta})$ is the likelihood function and $\text{Pr}(\bm{x})$ is the normalization constant, assuming that $\bm{\vartheta} \in \mathbb{R}^{C-1}$ for a given $\bm{x}\in \mathbb{R}^m$ follows a multivariate normal distribution $\bm{\vartheta}|\bm{x} \thicksim \mathcal{N}(\Theta^T\mathbf{x},\alpha^2\mathbf{I})$, $\Theta \in \mathbb{R}^{m\times C-1}$,$\text{Pr}(\bm{\vartheta}|\bm{x})$ is also multivariate normally distributed. Furthermore, the prior parameters are assumed to follow a normal distribution, i.e. $\bm{\theta} \sim \mathcal{N}(\mathbf{0},\beta ^2\mathbf{I})$ where $\beta$ is known. The parameter matrix $\mathbf{\Theta}$ is thus estimated by optimizing an equivalent of a regularized least squares function

\begin{align}\label{EstKMA}
\begin{split}
\mathbf{\hat{\Theta}}&=\underset{\Theta}{\arg\min} \mathbf{U(\Theta)}\\
U\mathbf{(\Theta)}&=\|\bm{\vartheta}-\mathbf{X\Theta}\|_F^2+\lambda\|\bm{\Theta}\|_F^2,
\end{split}
\end{align}
where $\bm{\vartheta}=[\bm{\vartheta}^{(i)}]_{i=1,2, \ldots,n}^T,\ \mathbf{X}=[\bm{x}^{(i)}]_{i=1,2, \ldots,n}^T,\ \|.\|_F$ is the Frobenius norm of a matrix and $\lambda$ is the regularization parameter. The result is a closed form estimate given by
$
\mathbf{\hat{\Theta}}=(\mathbf{X}^T\mathbf{X}+\lambda\mathbf{I})^{-1}\mathbf{X}^T\bm{\vartheta}. 
$
To capture non-linearities which may be present, a dual representation $\mathbf{\Theta}=\mathbf{X}^T\bm{\Gamma}$  is taken so that
\begin{align*}
U(\mathbf{\Gamma})= \|\bm{\vartheta}-\mathbf{X}\mathbf{X}^T\Gamma\|_F^2+\lambda\|\mathbf{X}^T\bm{\Gamma}\|_F^2
\end{align*}
then $U(\mathbf{\Gamma})$ is optimized to get $\hat{\bm{\Gamma}}=(\mathbf{K}+\lambda \mathbf{I})^{-1}\bm{\vartheta}$, where $\mathbf{K}=\mathbf{XX}^T$ is the Gram matrix, $K_{ij}= \langle \bm{x}^{(i)},\bm{x}^{(j)}\rangle + 1$. However a more general kernel $K_{ij}= ((\phi( \bm{x}^{(i)}),\phi(\bm{x}^{(j)}))$ where $\phi(\cdot)$is a nonlinear mapping, is preferred in practice.

The final step of the algorithm involves prediction/classification given a new set of response variables $\bm{x}^{new}$. This entails estimation of $\bm{\vartheta}^{new}$ by $\bm{\vartheta}^{new}=\hat{\mathbf{\Gamma}}^T\hat{\bm{x}}^{new}$, but $\hat{\bm{x}}^{new}=K((\phi( \bm{x}^{(i)}),\phi(\bm{x}^{(new)}))$. The computed $\bm{\vartheta}^{new}$ is used to estimate $\mathbf{t}^{new}$  by using $\bm{t}^{new}=\text{logit}^{-1} (\bm{\vartheta}^{new})$. The inverse of a logit function is given by 

\begin{align}\label{InvLogit}
\begin{split}
t_j^{new}&=\frac{\text{exp}\{\vartheta_j^{new}\}}{1+\sum\limits_{i=1}^{C-1}\text{exp}\{\vartheta_j^{new}\}},\quad j=\{1,2,\ldots,C-1\} \\
t_J^{new}&=\frac{1}{1+\sum\limits_{i=1}^{C-1}\text{exp}\{\vartheta_j^{new}\}}.
\end{split}
\end{align}

The class labels associated with $\bm{x}^{new}$ are then computed using the estimated conditional distribution by finding the components that maximize those of $\bm{t}^{new}$ i.e. using the Bayes rule. The computed $\mathbf{t}^{new}$ is then used to get the class label  ($\bm{\hat{y}}^{new}$) of the new data; for details see \cite{Dalmau2015}.

\section{Partial least squares (PLS) and some of its applications in genomics}
PLS is a very useful approach because it is able to analyze data with many, noisy, collinear as well as incomplete variables. 
PLS is usually utilized in data reduction when there is multicollinearity or when the data have more variables than the number of samples. Essentially, the PLS aims at maximizing the covariance between the response variables $\bm{Y}$ and the predictors $\bm{X}$, i.e., $ cov (\bm{X}^T\bm{Y})$ of highly multidimensional data by finding a linear subspace of the explanatory variables \citep {Wold2001, Hosk1988}. Some literature on PLS can be found in \cite{Wold2001,Wold1984,and Hosk1988}, among others.

The research on PLS is still very active due to its ability to address problems associated with the high dimensional data such as  multicollinearity and high dimensionality, among others. In the recent past, PLS has been utilized predominantly in high dimensional data in different fields like chemometrics and the ``omics'' like genomics, proteomics, metabolomics, and many other fields that generate large amounts of data, like spectroscopy \citep{Gromski2015}.  Recent applications of PLS in microarray studies include \cite{Huang2013}, who applied PLS regression (PLSR) in breast cancer intrinsic taxonomy, for classification of distinct molecular sub-types by using PAM50 signature genes as predictive variables in PLS analysis and the latent binary gene component analyzed by a logistic regression for each molecular sub-type. Also, \cite{Telaar2013} extended the notion of PLS-discriminant analysis (PLS-DA) to Powered PLS-DA (PPLS-DA), introducing a `power parameter' maximised towards the correlation between the components and the group-membership, thereby achieving a minimal classification error. Furthermore, \cite{Xi2014} discussed the PLS-DA with applications to metabolites data. Other articles involving the usage of PLS include: \cite{Dong2014} who used PLS to investigate the underlying mechanism of the post-traumatic stress disorder (PTSD) using microarray data; \cite{Gusnanto2013}, who made gene selection based on partial least squares and logistic regression random-effects (RE) in classification models; gene selection involving PLS was also done by \cite{Wang2015}. The sparse PLS has also been utilized by many researchers; for instance, \cite{Chun2009, Lee2011} and \cite{ Chung2010 } provided an efficient algorithm for the implementation of sparse PLS for variable selection in high dimensional data. Furthermore, \cite{Kim-Anh2008} used sparse PLS for variable selection when integrating omics data. They implemented sparsity via lasso penalization of the PLS loading vectors when computing the singular value decomposition.

\section{PLS generalized linear regression algorithm}
\label{Sec: PLSGLR}
In this section, we present an algorithm that can be applied to any Generalized Linear Regression which was developed by \cite{Bastien2005}. Consider the response data $\mathbf{y}$ with the explanatory variables $ \mathbf{x_1,\ldots,x_p}$; then a PLS-General Linear Regression (PLSGLR) can be written as 
\begin{align}
g(\theta)=\sum_{h=1}^{m}c_h\Big (\sum_{j=1}^{p}w_{hj}^{*}\mathbf{x}_j\Big) ,
\end{align}
where $\theta$ a conditional expectation of the variable $\mathbf{y}$ if its distribution is continuous, or a vector of probabilities if the variable $\mathbf{y}$ follows a discrete distribution with a finite support, while $g(.)$ is the link function chosen according to the probability distribution of $\mathbf{y}$. The PLS components are given by $ t_h=\sum_{j=1}^{p}w_{hj}^{*}\mathbf{x}_j, j=1, \ldots, p, h=1,\ldots,m $. To compute the PLS components, let $ \mathbf{X}= \mathbf{x_1 \ldots,x_p}$ be a matrix of $p$ centred explanatory variables $\mathbf{x}_j$'s. The key objective is to determine $m$ orthogonal PLS components defined as a linear combination of the $\mathbf{x}_j$'s. The algorithm is presented as follows:

\begin{enumerate} 
	\item Computation of the first PLS component $t_1$ :  First, the regression coefficients $a_{1j}$ of $\mathbf{x}_j$ are computed using the usual GLM procedure  of $\mathbf{y}$ on $\mathbf{x}_j, j=1\ldots p$. The column vector $\mathbf{a}_1$ which contains $a_{1j}$ is then normalized : $\mathbf{w}_1 =\mathbf{a}_1/\lVert \mathbf{a}_1\rVert $. Finally, the component $t_1$ is computed as  $\mathbf{t}_1=\mathbf{Xw}_1/\mathbf{w}_1' \mathbf{w}_1$. 
	
	\item Computation of the second PLS component $t_2$ : Involves the computation of the linear model coefficients $a_{2j}$ of $\mathbf{x}_j$ in the GLM setting of  $\mathbf{y}$ on $t_1$ and $\mathbf{x}_j, j=1,...,p$.  Since the main idea of PLS is to create the orthogonal components $t_2$, the component $t_1$ is added as a variable in estimating $\mathbf{y}$ on $t_1$ and $\mathbf{x}_j, j=1,...,p$. This is because the structure of PLSGLR does not allow the residuals of $y$ to be obtained in each iteration that would aid in construction of orthogonal components. The column vector $\mathbf{a}_2$ which contains $a_{2j}$ is normalized: $\mathbf{w}_2 =\mathbf{a}_2/\lVert \mathbf{a}_2\rVert $ and thereafter, the residual matrix $\mathbf{X}_1$ is obtained via the regression of $\mathbf{X}$ on $t_1$. The use of residual matrix in the attainment of the next component ensures orthogonality between the different components. The component $t_2$ is calculated by  $\mathbf{t}_2=\mathbf{X}_1\mathbf{w}_2/\mathbf{w}_2' \mathbf{w}_2$.  Finally, $\mathbf{t}_2$ is expressed in terms of $\mathbf{X}$ : $\mathbf{t}_2=\mathbf{Xw}_2^*$. 
	
	\item Computation of the  $h^{th}$ PLS Component $t_h$: Consider  the already computed components   $t_1,...,t_{h-1}$; the final component $t_h$ is computed by calculating the GLM coefficients $a_{hj}$ of $\mathbf{x}_j$ by fitting  $\mathbf{y}$ on $t_1,...t_{h-1}$ and $\mathbf{x}_j, j=1,...,p$. Next, the column vector $\mathbf{a}_h$, which contains $a_{hj}$ is normalized as: $\mathbf{w}_h =\mathbf{a}_h/\lVert \mathbf{a}_h\rVert $.  The residual matrix $\mathbf{X}_{h-1}$ of the regression of $\mathbf{X}$ on $t_1,...,t_{h-1}$ is then computed.  The use of the residual matrix and the previously obtained $t_1,...t_{h-1}$  as covariables in calculating the GLM coefficients helps the creation of orthogonal components, as previously explained. The final component $t_h$ is thus computed as $\mathbf{t}_h=\mathbf{X}_{h-1}\mathbf{w}_h/\mathbf{w}_h' \mathbf{w}_h$. Finally, $\mathbf{t}_h$ is expressed in terms of $\mathbf{X}$ : $\mathbf{t}_h=\mathbf{Xw}_h^*$.
	
\end{enumerate}

\cite{Bastien2005} note that while computing the components $t_h$, the nonsignificant elements in $a_h$ can be set to zero in order to simplify calculations, since only the significant response variables are needed to build the PLS components. The number of $m$ components to be used can be determined through cross-validation or by hard thresholding. The iteration can be stopped once there are no more significant coefficients in $a_h$. 

Consider $x_{h-1,i}$, a column vector of the transpose of the $i$th row of $X_{h-1}$; then $t_{hi}=x'_{h-1,i}w_h/w'_hw_h$ of the $i$th case on the component $t_h$. This is basically the slope of the fitted line of the univariate OLS linear regression without intercept for $x_{h-1,i}$ on $w_h$, which can be estimated even with some data missing. Consequently, the component is computed based on the available data. Therefore the PLSGLR algorithm by \cite{Bastien2005} effectively copes up with missing data.  

\section{Applications to real data sets}
\label{chap:AppRealData}
\subsection{Some exploratory analysis}
In this study, we will describe in detail the analysis of the Colon data by \cite{Alon1999}, obtained from the \verb|R| package \verb|plsgenomics|, which consists of a $(62 \times 2000)$ matrix giving the expression levels of 2000 genes for 62 colon tissue samples.

\begin{figure}[H]
	\includegraphics[height=245pt, width=430pt]{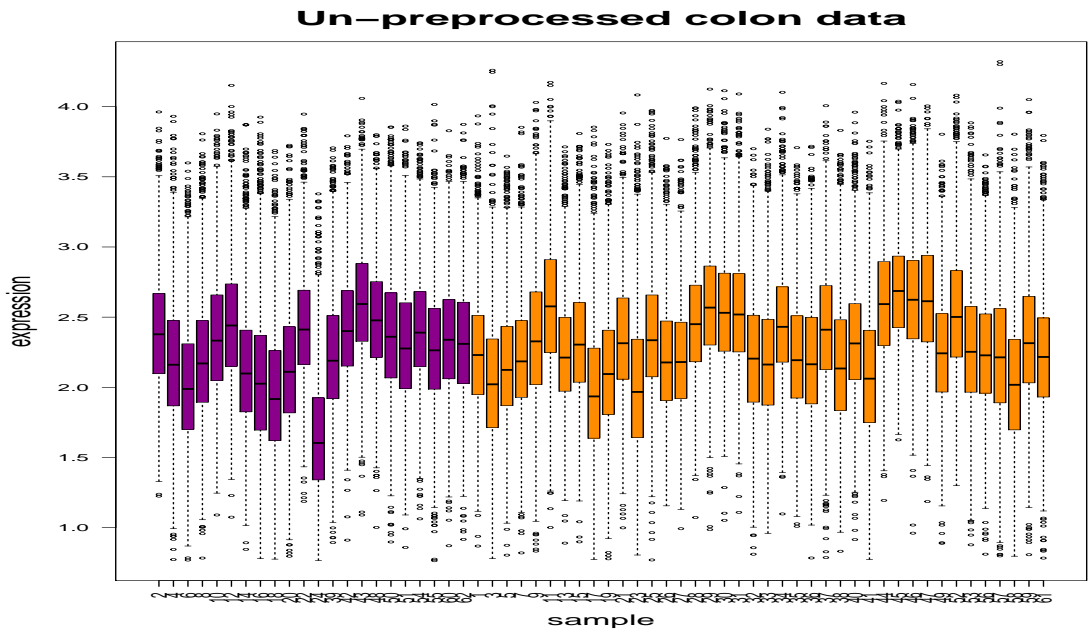}
	\caption{\textbf{Box plot for the un-preprocessed colon data}. The box plot for un-preprocessed data clearly shows that the data are noisy and have a lot of variations. The data have some unwanted variations that are expected to affect the analysis. They also lack symmetry.}
	\label{BoxColon1}
\end{figure}

\begin{figure}[H]
	\includegraphics[height=245pt, width=430pt]{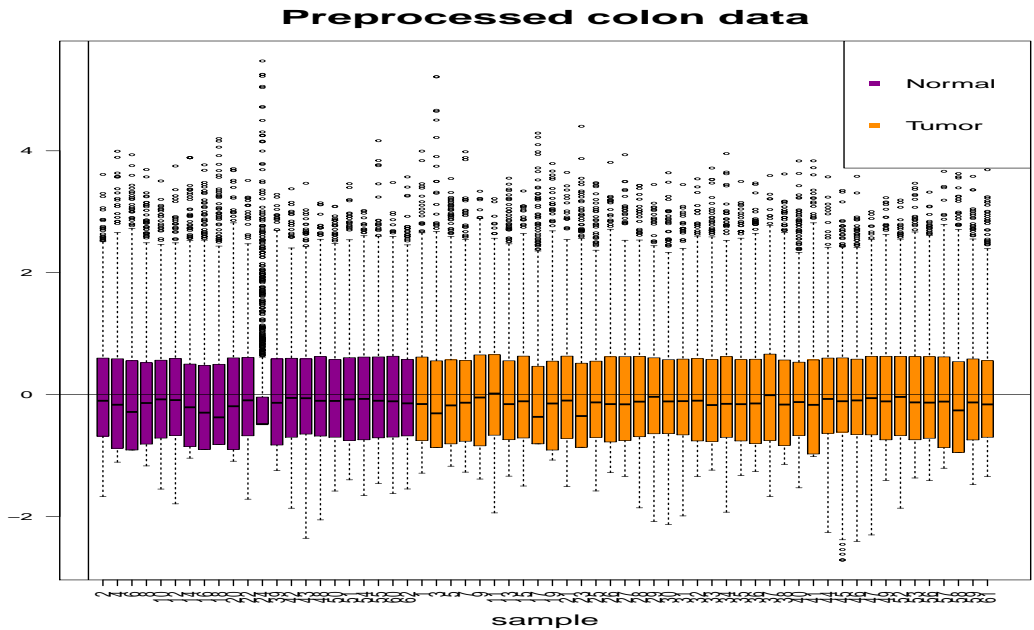}
	
	\caption{\textbf{Box plot for the preprocessed colon data}. This plot presents less variations. The data seem to have a symmetric distribution and do not show the presence of unwanted variation. From the two figures, it is expected that the preprocessed data would be easier to analyze.}
	\label{BoxColon2}
\end{figure}

\begin{figure}[H]
	\includegraphics[height=245pt, width=480pt]{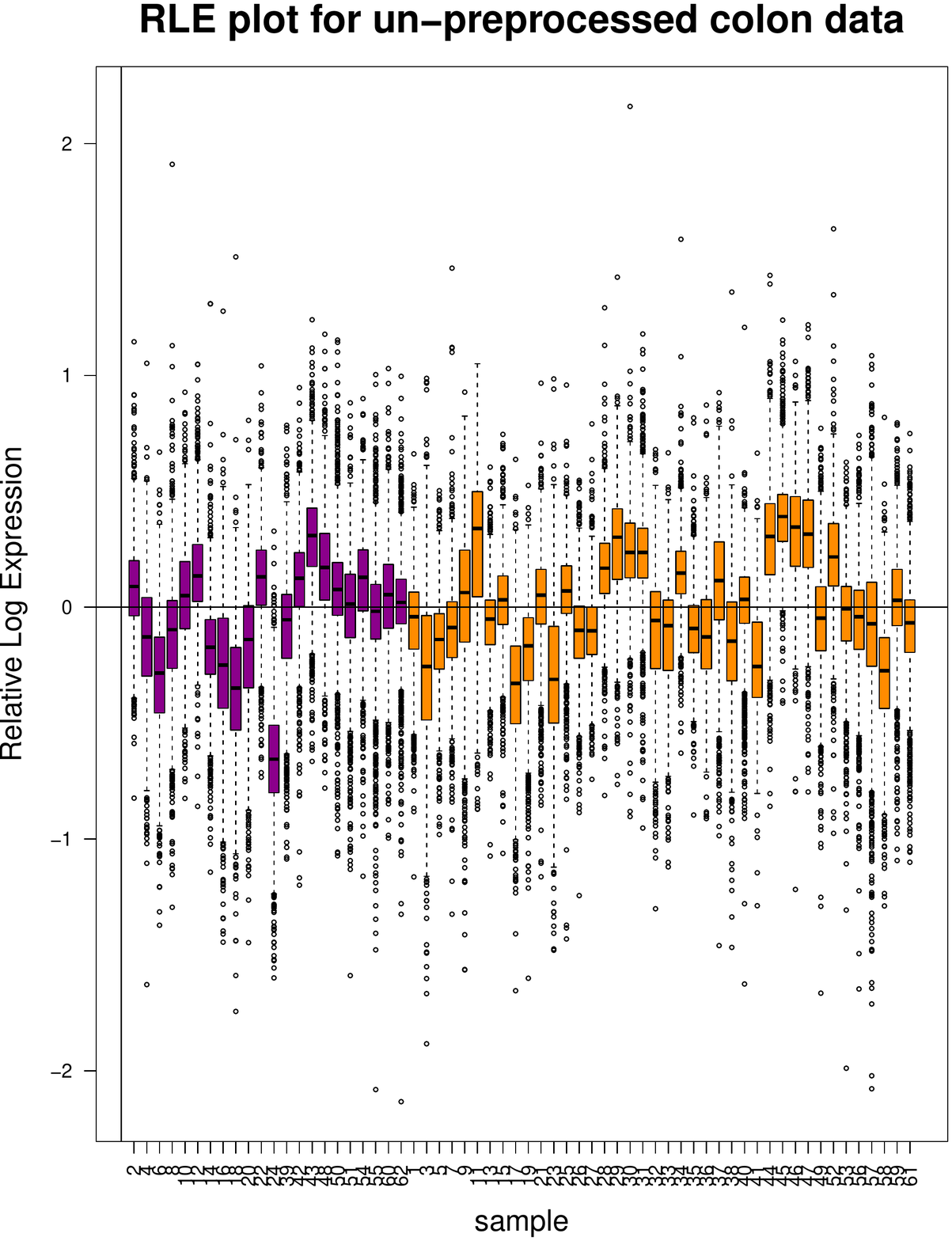}
	
	\includegraphics[height=245pt, width=480pt]{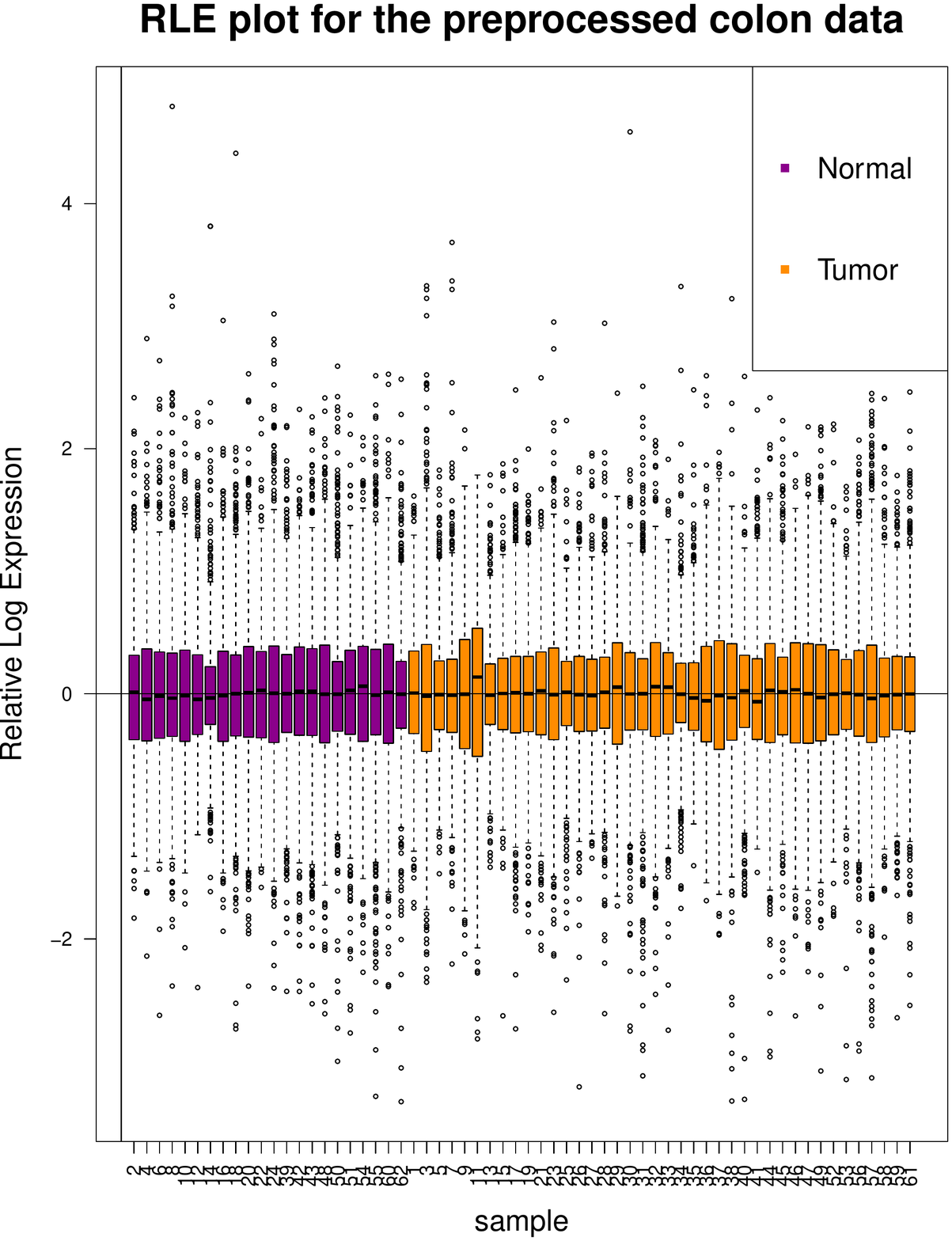}
	
	\caption{\textbf{RLE plots for the un-preprocessed and preprocessed colon data.} The RLE plot for the un-preprocessed data shows the presence of a lot of heterogeneity, implying that the data have variations that do not necessarily come from biological factors. However, the RLE plot for the processed data shows homogeneity and lack of unwanted noise, and should give better results when analyzed statistically.}
	\label{RLEcolon}
\end{figure}

\begin{figure}[H]
	\includegraphics[height=230pt, width=490pt]{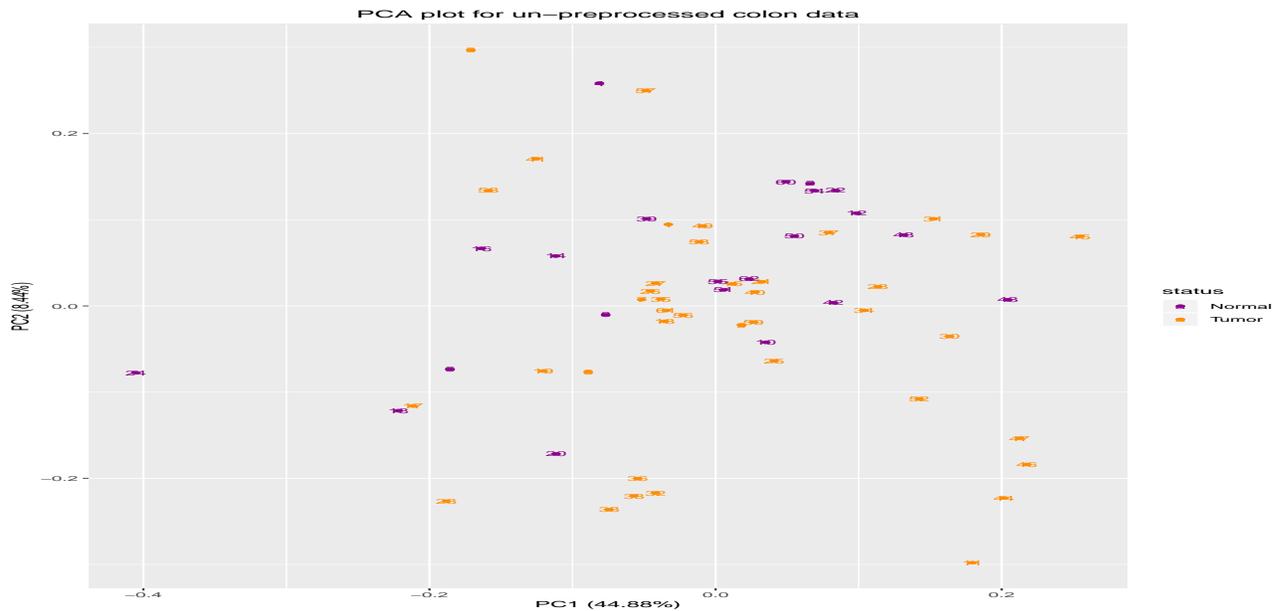}
	\caption{\textbf{PCA plot for the un-preprocessed Colon data.} The PCA plots show that it is harder to separate/classify the un-preprocessed data.}
	\label{PCAcolon1}
\end{figure}

\begin{figure}[H]
	
	\includegraphics[height=230pt, width=490pt]{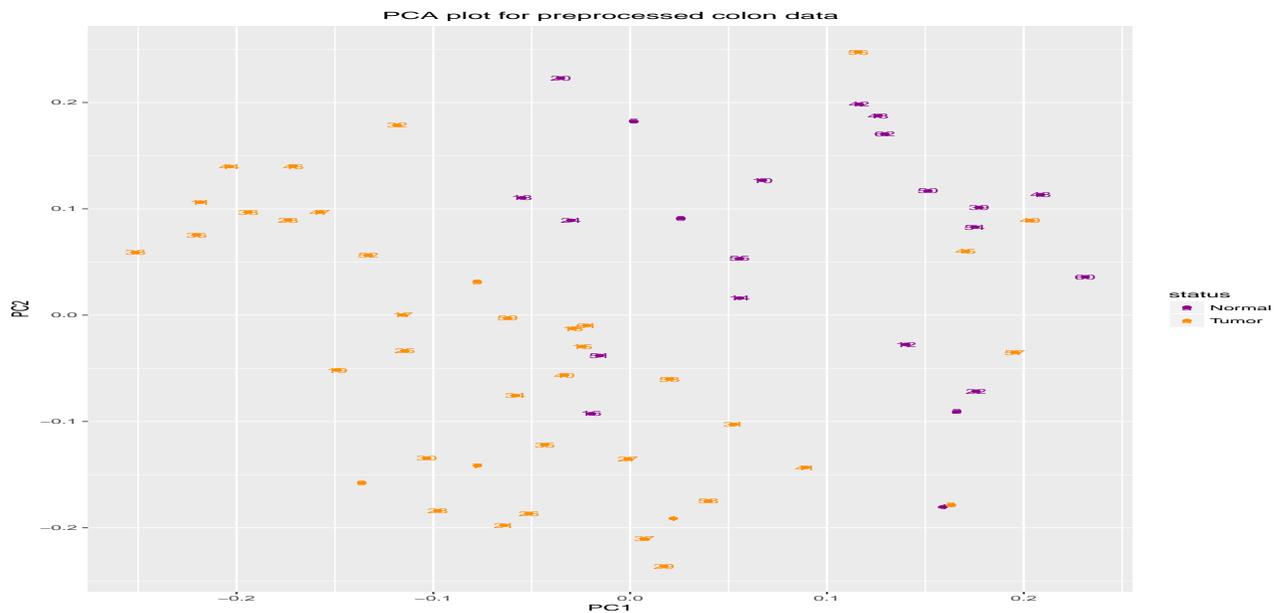}
	
	\caption{\textbf{PCA plots for the preprocessed Colon data.} It is relatively easier to separate/classify preprocessed data.}
	\label{PCAcolon2}
\end{figure}

An exploratory analysis of the data was done in order to visualize the differences in the un-preprocessed and preprocessed microarray data sets. The preprocessing is done using the \verb|R|  package \verb|plsgenomics| see \url{https://rdrr.io/cran/plsgenomics/}, that implements the recommendations of \citep{Dudoitetal2002}. To visualize the differences between the preprocessed and un-preprocessed data sets, we consider the pairs of box plots, relative log expression (RLE), and principal components analysis (PCA) plots presented in Figures \ref{BoxColon1},\ref{BoxColon2}, \ref{RLEcolon}, \ref{PCAcolon1} and \ref{PCAcolon2} respectively.


The same pair of data sets is examined using RLE plots, to show how the preprocessed data compares with the un-preprocessed data set with regard to the batch effect or any other abnormality. The RLE plots have been extensively used in studies of microarray data to reveal the effectiveness of data normalization; for an example see \cite{GagnonBartsch2011}. The RLE plots are simple yet very powerful in the visualization of data to detect unwanted variations. To understand how an RLE plot is constructed, consider a data matrix $\bm{X}_{p\times n}$ where $p$ is the number of genes while $n$ the number of microarray samples, and so the element of the data matrix $x_{ij}$ represents the $i^{th}$ gene in the $j^{th}$sample. The RLE plot is then constructed by first calculating the median across each of the $p$ rows, and then substracting the respective median across each row of the data matrix $\bm{X}$, i.e $(x_{ij}-\text{median}\  x_{i*})$. The median is used because it is robust and not affected by outliers. A box plot is then generated for each of the $n$ samples, and a good one will be centered around zero and its width (interquartile range) should be equal to or less than $0.2$ (see \citep{GagnonBartsch2011}).

\newpage
Finally we compare the ease of classification between the un-preprocessed and preprocessed data. The simplest way to visualize the separability of categories in a given data set is the use of principal components analysis (PCA) plots. 


\cite{GagnonBartsch2011} note that one of the key challenges of the removal of unwanted variation is the difficulty in distinguishing the unwanted variations from the biological variation of interest. Furthermore, they note that the most appropriate way to deal with unwanted variation depends on the final objective of the analysis, for instance: differential expression (DE), classification, or clustering.

\subsection{Analysis of the un-preprocessed data}
In this analysis, we compare the performance of our proposed model extensions PLSGLR-log, PLSGLRDA and the KMA \cite{Dalmau2015} to that of the classical methods when the data has neither been preprocessed nor variables been selected, thus testing the performance of the classification algorithms in the presence of noise. The performance of the methodologies is then compared using a 10 fold cross validation (10-CV) and the corresponding classification error rates are computed. The results are presented in Table \ref{CrossVal}.

\begin{table}[H]
	\caption{Rate of classification error for the different methods when applied to the un-preprocessed data set}
	\vspace{.5cm}
	\label{CrossVal}
		\begin{tabular}{lcccccccc}
			\hline				
			DATA   & PLSGLR-log & PLSGLRDA & KNN & LDA & PLSDA & RPLS & SVM & KMA\\
			\hline
			Colon  &38.3& 31.7& 60.0& 25.0 &11.7& 15.0& 18.3&1.7 \\ 
						\hline    
	\end{tabular}
\end{table}

A particular method is judged to be the ``best'' if it has a lower classification error rate relative to the other methods, otherwise it is a poor classifier. The results based on minimal cross validation classification error rates indicate that for the Colon data, the KMA emerges as the best, followed by PLSDA, and RPLS, while the worst were KNN and PLSGLR-log. 

\subsection{Analysis of preprocessed data}
\label{Sec:AnalysPrepData}
During the preprocessing of microarray data the feature selection step is usually performed. This is because out of the thousands of variables (genes) generated, only a handful may play an important role towards the biological problem of interest. The thousands of data points are likely to be noisy due to biological or technical reasons. Thus the feature selection  extracts a  subset of the genes that are most informative (optimum subset of features). This reduces the noise  by removing irrelevant or redundant features \citep{Awada2012,Dudoitetal2002}. Most commonly used feature  selection methods involve ranking the genes based on some value of a univariate statistic, like the $t$-statistic, the F-statistic, or the Wilcoxon and Kruskal-Wallis statistics. A cut-off point based on either the number of genes or the p-value is imposed, to determine the number of variables to be used. \cite{Dudoitetal2002} suggest a gene selection method based on ranking. This is achieved by finding the ratio of between-group  to within-group sum of squares $(BSS/WSS)$ so that for a gene $j$,
\begin{equation}\label{FS}
BSS_j/WSS_j=\frac{\sum_{i}^{} \sum_{k}^{}I(y_i=k)(\bar{x}_{kj}-\bar{x}_{.j})^2}{\sum_{i} \sum_{k}I(y_i=k)(x_{ij}-\bar{x}_{kj})^2}
\end{equation}

\noindent where $\bar{x}_{.j}$ and $\bar{x}_{kj}$ are the average expression levels of gene $j$ and across all samples in class $k$, respectively. The $p$ genes with the biggest ratio are selected. In this study, we adopted the \cite{Dudoitetal2002} method of feature selection.

The preprocessing and the gene selection were performed using the recommendations of \cite{Dudoitetal2002}. The top $p$ genes were thus selected using Equation \ref{FS} for the implementation of the classification methods.

The classification error rates for the various methodologies when applied to the data under consideration are presented in Table \ref{PrepData}.

\begin{table}[H]
	\caption{ classification error rates for the different methods when applied to the preprocessed data set}
	\vspace{.5cm}
	\label{PrepData}
		\begin{tabular}{lcccccccc}
			\hline				
			DATA   & PLSGLR-log & PLSGLRDA & KNN & LDA & PLSDA & RPLS & SVM & KMA\\
			\hline
			Colon  &16.4& 13.3& 26.7&15.0 &11.7&11.7 &14.8 &11.2\\ 
			\hline    
	\end{tabular}
\end{table}

The results indicate that KMA was the best, followed by RPLS, PLSDA. PLSGRDA performed equally well, while KNN emerged as the worst classifier, also in every comparison. 

\section{Summary and conclusions}
In this study, two extensions of the PLSGLR were considered in addition to the KMA for a comparative study with some classical classification methodologies, namely KNN, LDA, PLSDA, RPLS and SVM, when applied to one commonly used microarray data set. The data were considered when un-preprocessed and when preprocessed. For both the un-preprocessed and preprocessed cases, the KMA emerged as a clear ``winner"  based on lower classification error rates. The KMA algorithm can therefore be recommended for  classification problems involving noisy and non-noisy data. This could be due to the fact that the chosen kernels map the samples to a higher dimensional space, where they become linearly separable. This leads to a better classification ability by the KMA. Furthermore, the three new algorithms can therefore be considered as an addition to the existing literature for the microarray data classification problems.

\section*{Acknowledgements}
We acknowledge the partial support from the Mexico's Consejo Nacional de Ciencias y Tecnolog\'ia (CONACyT) project number 252996.

\bibliographystyle{cbe}
\bibliography{PLSGLR_REF}

\end{document}